\documentclass[useAMS,usenatbib]{mn2e}
\usepackage{graphicx} 
\usepackage{color}
\usepackage{float}
\usepackage{placeins}
\usepackage{multicol}
\usepackage{amsmath}
\usepackage{verbatim}
\usepackage[justification=centering]{caption}
\usepackage{array}
\usepackage{booktabs}
\usepackage{threeparttable}
\usepackage{amssymb}% http://ctan.org/pkg/amssymb
\usepackage{pifont}% http://ctan.org/pkg/pifont
\usepackage{array,multirow}
\def\be{\begin{equation}} 
\def\ee{\end{equation}}

\def\msun{{\Msun}}

\def\gsim{\lower.5ex\hbox{\gtsima}} 
\def\lsim{\lower.5ex\hbox{\ltsima}} \def\gtsima{$\; \buildrel > \over 
\sim \;$} \def\ltsima{$\; \buildrel < \over \sim \;$} \def\prosima{$\; 
\buildrel \propto \over \sim \;$} \def\gsim{\lower.5ex\hbox{\gtsima}} 
\def\lsim{\lower.5ex\hbox{\ltsima}} 
\def\simgt{\lower.5ex\hbox{\gtsima}} 
\def\simlt{\lower.5ex\hbox{\ltsima}} 
\def\simpr{\lower.5ex\hbox{\prosima}} \def\la{\lsim} \def\ga{\gsim} 
  
 \def\gtsima{$\; \buildrel > \over \sim \;$} 
\def\ltsima{$\; \buildrel < \over \sim \;$} 
\def\gsim{\lower.5ex\hbox{\gtsima}} 
\def\lsim{\lower.5ex\hbox{\ltsima}} 
\def\simgt{\lower.5ex\hbox{\gtsima}} 
\def\simlt{\lower.5ex\hbox{\ltsima}} 
\def\simpr{\lower.5ex\hbox{\prosima}} 
\def\la{\lsim} 
\def\ga{\gsim}

\def\msun{\,{\rm \Msun}}

\def\E3{{\cal E}_{\rm g}^{III}}

\def\Msun{\rm M_\odot}

\def\Msun{\rm M_\odot}

\def\myr{\rm Myr }

\def\M*{M_*}
\def\Z*{Z_*}
\def\L*{L_*}

\title[Mass assembly of early black holes]{The mass assembly of high-redshift black holes} 
\author[Piana et al.]{Olmo Piana$^{1}$\thanks{piana@astro.rug.nl}, Pratika Dayal$^1$, Marta Volonteri$^2$, Tirthankar Roy Choudhury$^3$\\ 
\\
$^{1}$ Kapteyn Astronomical Institute, University of Groningen, P.O. Box 800, 9700 AV Groningen, The Netherlands \\
$^{2}$ Sorbonne Universites, UPMC Univ Paris 6 et CNRS, UMR 7095, Institut d'Astrophysique de Paris, 98 bis bd Arago, 75014 Paris, France \\ 
$^{3}$National Centre for Radio Astrophysics, Tata Institute of Fundamental Research, Pune 411007, India\\
}

\begin{document} 
 
\date{} 

\maketitle

\begin{abstract}
We use the \textit{Delphi} semi-analytic model to study the mass assembly and properties of high-redshift ($z>4$) black holes over a wide mass range, $10^3 < M_{bh}/\msun < 10^{10}$. Our black hole growth implementation includes a critical halo mass ($M_{h}^{crit}$) below which the black hole is starved and above which it is allowed to grow either at the Eddington limit or proportionally to the gas content of the galaxy. As a consequence, after an initial growth phase dominated by black hole mergers down to $z \sim 7 ~ (9)$, supermassive black holes in $z=4$ halo masses of $M_h|_{z=4} \sim 10^{11.75} ~ (10^{13.4}) \msun$ mainly grow by gas accretion from the interstellar medium. In particular, we find that: {\it (i)} while most of the accretion occurs in the major branch for $M_h|_{z=4} \sim 10^{11-12} \msun$ halos, accretion in secondary branches plays a significant role in assembling the black hole mass in higher-mass halos ($M_h|_{z=4} \gtrsim 10^{12} \msun$); {\it(ii)} while the Eddington ratio increases with decreasing redshift for low-mass ($M_{bh} < 10^5 \msun$) black holes, it shows the opposite trend for larger masses. In addition, since the accretion rate depends on the gas mass present in the host halo, the duty cycle of the Eddington-limited accretion phase -- which can last up to $\approx 650$ Myr -- is crucially linked to the joint assembly history of the black hole and its host halo.
\end{abstract}


%#################################################################
\section{Introduction}
%#################################################################

High-redshift observations have discovered that about 200 quasars, powered by accretion onto supermassive black holes (SMBH), were already in place in the first billion years of the Universe \citep[e.g.][]{fan2001, mortlock2011, mazzucchelli2017, banados2018, matsuoka2018}. The masses of these black holes can be as high as $M_{bh} \sim 1.2 \times 10^{10} \Msun$ at a redshift of $z =6.3$ \citep{wu2015} and $M_{bh} \sim 7.8 \times 10^8 \Msun$ at $z=7.54$ \citep{banados2018}. The discoveries of such early SMBHs have re-ignited the - still largely unanswered - question of how they could have grown so large so fast \citep[for a recent review see][]{inayoshi2019}. Growing a black hole of $10^9 \Msun$ from a seed of about $100 \Msun$ (at a reasonable radiative efficiency of $10\%$) would require near-Eddington accretion for almost 800 Myr - this corresponds to the age of the universe at $z \sim 6.8$ \citep{haiman2001, tanaka2014}. The problem is exacerbated by the fact that high-velocity recoil kicks (of up to $500-1000 \mathrm{km s^{-1}}$) can significantly reduce the efficiency of black hole growth through mergers \citep{fitchett1983, favata2004, haiman2004, volonteri2006} by expelling them from the shallow potential wells of low-mass halos. While some studies have tried to alleviate this problem by allowing black holes to undergo short episodes of super-Eddington accretion within the context of a slim accretion disk model \citep{haiman2004, yoo2004, volonteri2005, lupi2015, pezzulli2016, mayer2018}, others have suggested that pristine gas clouds could collapse isothermally and without fragmentation to yield direct-collapse black hole (DCBH) seeds with mass $M_{bh} \sim 10^4-10^5 \msun$ \citep{haehnelt1993, loeb1994, eisenstein1995, bromm2003, begelman2006, habouzit2016, luo2018}. Alternatively, \cite{boco2020} recently proposed that heavy black hole seeds of masses $10^4 - 10^6 \msun$ could form via merging of compact stellar remnants in young galaxies with dense gas environments.

A huge amount of theoretical effort has been dedicated to modelling black hole evolution at high redshift using both semi-analytic models \citep[SAMs; e.g.][]{volonteri2008, croton2006, bower2006, hirschmann2012, barausse2012, sesana2014, mutch2016, lacey2016, valiante2016, marshall2019} and hydrodynamic simulations \citep[e.g.][]{schaye2015, khandai2015, dimatteo2017, dubois2014, volonteri2016b, vogelsberger2014, sijacki2015, latif2018, habouzit2019, huang2019}. While SAMs offer speed and flexibility in exploring a larger parameter space and uncovering the properties of the black hole population as a whole, simulations offer the advantage of tracking the dynamics of the different galactic components. Nevertheless, zoom-in simulations are required both to reach the parsec-scale resolution needed to accurately model black hole accretion during the early growth phase \citep{lupi2019}, when the internal structure of the host gas cloud plays a major role \citep{beckmann2019}, and to follow the (sub-) kiloparsec-scale gas inflows from the cosmic web which drive black hole growth at later evolutionary stages \citep{costa2014, richardson2016}; these techniques, though, do not allow a statistical study of the black hole population as a whole.

In general these works are tuned to reproduce the low-redshift black hole--host galaxy data. Episodes of fast black hole growth are triggered by galaxy mergers, disk instabilities or bursts of star formation, with the black hole accretion rate often assumed to be equal to the Bondi-Hoyle rate and capped at one or two times the Eddington limit. Since the efficiency of black hole growth through mergers, which are much more common at high redshift, is hampered by the black hole merging timescale \citep{tremmel2018} and by recoil kicks \citep{haiman2004}, most works agree that SMBHs at $z>6$ mainly grow by gas accretion \citep[see][for a review]{wise2019}. Despite this progress in understanding the evolution of the high-redshift black hole population, many questions still remain open.

In this paper we want to study how the black hole physical properties and growth histories vary as a function of host halo mass across the entire population. To this end, we use the semi-analytic model {\it Delphi}, that has been shown to reproduce all key observables for galaxies and active galactic nuclei (AGN) at $z \gsim 5$ \citep[including the AGN and stellar ultra-violet luminosity function, the stellar mass function, the stellar mass density and the black hole mass function,][]{dayal2018} in addition to reionization observables \citep[the electron scattering optical depth and emissivity constraints,][]{dayal2020}. The key strengths of this model lie in that: (i) it is seeded with two types of black hole seeds (stellar and direct collapse); (ii) the black hole accretion rate is linked to the host halo mass; and (iii) it uses a minimal set of free parameters to model star formation, black hole accretion and their associated feedback. In particular, we make use of the results from numerical simulations \citep{dubois2015, bower2017, habouzit2017} finding that black hole growth is suppressed in low-mass halos, due to the fact that SN feedback in low-mass galaxies hinders the accumulation of gas mass around the central regions. With this picture in mind, we assume that black holes can not accrete below a critical halo mass threshold $M_h^{crit}$. So far, \textit{Delphi} is the only SAM explicitly implementing such a mechanism to regulate BH growth.

The cosmological parameters used in this work are $\Omega_{\rm m }, \Omega_{\Lambda}, \Omega_{\rm b}, h, n_s, \sigma_8 = 0.3089, 0.6911, 0.049, 0.67, 0.96, 0.81$ \citep{planck2015}. We quote all quantities in comoving units, unless stated otherwise, and express all magnitudes in the standard AB system \citep{oke1983}.

We start by describing the theoretical model in Sec. \ref{sec_model}. We then discuss the assembly histories of black holes and their host halos in Sec. \ref{sec_mass} and present the resulting key properties of black holes (occupation fraction, duty cycle and the mass function) in Sec. \ref{sec_res} before concluding in Sec. \ref{conclusions}.

%#################################################################
\section{Theoretical model}
\label{sec_model}
%#################################################################

%*******************************************************************
\begin{table*}\begin{center}
\begin{threeparttable}
\setlength{\extrarowheight}{3pt}
\begin{tabular}{ccccc}
\hline
\hline
Parameter & & Description & $ins1$ & $tdf4$\\
\hline
$\epsilon_r$ & & radiative efficiency of black hole accretion & 0.1 & 0.1\\
$f_*$ & & star formation efficiency threshold & 0.02 & 0.02\\
$f_*^w$ & & fraction of SN energy that couples to the gas & 0.1 & 0.1\\
$f_{bh}^w$ & & fraction of AGN energy that couples to the gas & 0.003 & 0.003\\
$f_{bh}^{ac}$ & & fraction of available gas mass that black holes can accrete & $5.5 \times 10^{-4}$ & $5.5 \times 10^{-4}$\\
$f_{Edd} (M_h < M_{h}^{crit})$ & & black hole accretion rate in fraction of Eddington & $7.5 \times 10^{-5}$ & $7.5 \times 10^{-5}$\\
$f_{Edd} (M_h > M_{h}^{crit})$ & & black hole accretion rate in fraction of Eddington & 1 & 1\\
$\alpha$ & & LW background threshold for DCBH formation (in units of $J_{21}$) & 30 & 300\\
Reionization feedback & & - & No & Yes\\
Delayed mergers & & dynamical friction acting to delay the merging of the baryonic components & No & Yes\\
\noalign{\smallskip}
\hline
\hline
\end{tabular}
\caption{Here we list the free parameters of the model in column 1, we describe them in column 2 and we show their values in the $ins1$ and the $tdf4$ scenarios in columns 3 and 4, respectively.}
\label{table_models}
\end{threeparttable}
\end{center}\end{table*}
%*******************************************************************

In this work, we use the semi-analytic code \textit{Delphi} \citep[\textbf{D}ark matter and the \textbf{e}mergence of ga\textbf{l}axies in the e\textbf{p}oc\textbf{h} of re\textbf{i}onization;][]{dayal2014,dayal2018}, which jointly tracks the assembly of the dark matter, baryonic and black hole components of high-redshift ($z \gsim 4$) galaxies. In brief, starting at $z=4$, we build analytic merger trees for 550 halos, equally separated in log space in the mass range $M_h = 10^8-10^{13.5} \Msun$, up to $z=20$ in time steps of 20 Myr. Each $z=4$ halo is assigned a number density according to the Sheth-Tormen halo mass function \citep[HMF,][]{sheth2002} at $z=4$. We then assign the same number density also to all the progenitors of the halo; we have confirmed that the resulting HMFs are in accord with the Sheth-Tormen HMF at all redshifts for $z \sim 5-20$. 

The very first progenitors of each tree are seeded with an initial gas mass proportional to the halo mass such that $M_{gi} = (\Omega_b/\Omega_m) M_h$. Such first halos irradiated by a Lyman-Werner (LW) background of intensity ($J_{LW}$) larger than a critical value $J_{crit} = \alpha J_{21}$ (where $J_{21} = 10^{-21} \mathrm{erg s^{-1} Hz^{-1} cm^{-2} sr^{-1}}$ and $\alpha$ is a free parameter) are assigned a DCBH seed with a mass between $10^3-10^4 \Msun$. First halos (at $z>13$) not fulfilling this criteria are instead assigned a stellar black hole (SBH) seed with a mass of $150 \msun$. Starting from these first progenitors and going forward in time, at each redshift step the halos and their baryonic components can grow both through the smooth accretion of dark matter and gas from the intergalactic medium (IGM) and through mergers; similarly, black holes grow through accretion from the interstellar medium (ISM) and mergers. 

At a given time, the star formation efficiency is defined as the minimum between the type-II Supernovae (SNII) energy required to unbind the rest of the gas and an upper threshold $f_*$. In the interest of simplicity, each newly formed stellar population follows a Salpeter initial mass function \citep{salpeter1955} over $0.1-100 \Msun$, has a metallicity $Z = 0.05\ Z_\odot$ and an age of $2\ \myr$. A fraction of the gas mass $M_{*}^{gf}(z)$ left in the halo after the star formation and SNII feedback can be accreted onto the black hole. The black hole accretion rate depends on the mass and the redshift of the host halo through a critical halo mass $M_{h}^{crit}(z) = 10^{11.25}[\Omega_m(1+z)^3 + \Omega_\lambda]^{-0.125}$, below which the black hole is stuck in a stunted accretion mode. Once the halo outgrows $M_h^{crit}$, at each time step its black hole is allowed to accrete either a fixed fraction $f_{bh}^{ac}$ of the gas mass present in the galaxy or a mass corresponding to what it would accrete if it were growing at the Eddington rate, whichever lower. The physical reasoning behind $M_h^{crit}$ is the following: while gas inflows towards the central black hole are disrupted by SN feedback in low-mass galaxies \citep[see for instance][]{dubois2015, habouzit2017, bower2017, angles2017}, \cite{lupi2019} have shown that the black hole mass growth speeds up at later times, once the host galaxy reaches a stellar mass of $M_* \approx 10^{10} \msun$, and SN feedback is not effective in keeping the gas away from the central regions anymore. In our work, the transitional host mass $M_h^{crit}(z)$ has been tuned to simultaneously reproduce the number densities of low-luminosity AGN at $z \geq 5$ and the black hole mass function at $z=6$ \citep[see][]{dayal2020}.

The mass accreted by the black hole at any redshift step can then be written as
\begin{equation}
M_{bh}^{ac}(z) = \min\left[f_{Edd}M_{Edd}(z),\ (1-\epsilon_r)f_{bh}^{ac}M_{*}^{gf}(z) \right],
\label{eq_condition}
\end{equation}
where $M_{Edd}(z) = \dot{M}_{Edd} \times \Delta t$ is the mass accreted by the black hole in a time step assuming the Eddington growth rate, $f_{bh}^{ac} = 5.5 \times 10^{-4}$ represents the maximum fraction of the total gas mass left in the host galaxy that can be accreted onto the black hole, $\epsilon_r = 0.1$ is the radiative efficiency of the black hole and $f_{Edd}$ is defined as 
\begin{equation}
f_{Edd} = \begin{cases} 
      7.5 \times 10^{-5} & M_h(z) < M_{h}^{crit}(z) \\
      1 & M_h(z) \geq M_{h}^{crit}(z) \\
   \end{cases}
   \label{eq_fedd}
\end{equation}
A fixed fraction $f_{bh}^w = 3 \times 10^{-3}$ of the total energy emitted by the accreting black hole is allowed to couple to the gas content of the host halo.

While we assume the mergers between dark matter halos to occur instantaneously, we explore two prescriptions for the mergers of their baryonic components: instantaneous, and merging after a delay induced by dynamical friction \citep{lacey1993}. Finally, we include the impact of reionization feedback. Creating an ultra-violet (UV) heating background, reionization can photo-evaporate the gas mass from low-mass halos \citep{dayal2018}. We implement this feedback in its maximal form by suppressing the gas content of all halos with a virial velocity $V_{vir}(z) < 40\ \mathrm{km s^{-1}}$.

To test the robustness of our approximations, we present results for two cases: (i) the {\it instantaneous} model ($ins1$) corresponds to galaxies and black holes merging instantaneously with their host halos, no UV feedback and DCBH seeds using a value of $\alpha=30$; (ii) the {\it delayed} model ($tdf4$) includes the prescription for the dynamical friction delay, a higher value of $\alpha=300$ for DCBH seed formation (which effectively reduces the number of DCBH seeds by a factor of $\approx 50$) and the impact of the reionization feedback. These models and their free parameter values used are summarised in In Table~\ref{table_models}. 

We now briefly discuss the interplay between the baryonic processes implemented in our model and black hole growth. Gas accretion onto the host galaxy (either from the IGM or through galaxy mergers), star formation, SN feedback and AGN feedback all contribute in regulating the amount of gas mass in the galaxies, thereby having a direct effect on black hole growth. The extent of their impact changes across the different growth phases of the black hole and its host. Increasing the star formation rate threshold $f_*$ or the coupling factor $f_*^w$ means that SN feedback are more effective in ejecting gas, especially in low-mass galaxies. These parameters are tuned to match the galaxy UV luminosity function, the stellar mass function, the stellar mass density and the star formation rate density. As the host halo grows and the potential well deepens, though, SN feedback becomes less effective. The value of $f_{Edd}(M_h < M_h^{crit})$ (the fraction of the Eddington rate at which the black hole is allowed to accrete), is tuned to match the faint end of the AGN UV luminosity function. Once $M_h > M_h^{crit}$ and the Eddington-limited accretion phase has started, the impact of AGN feedback grows, along with the black hole mass, with an exponential trend. We tested the values of the parameters strictly related to black hole growth ($M_h^{crit}$, $f_{bh}^{ac}$ and $f_{bh}^w$) by looking at how changing them would affect the AGN luminosity function and the black hole mass function. An increase in $M_h^{crit}$ leads to a decrease of the amplitude of the black hole mass function at $M_{bh} \gsim 10^4 \msun$, to a lower normalisation of the AGN luminosity function and to a slightly higher number density of very luminous AGN. In fact a higher $M_h^{crit}$ means a higher gas mass at the onset of the Eddington-limited accretion phase, which can then be sustained for a longer period of time. The parameter $f_{bh}^{ac}$ directly intervenes in Eq.~\ref{eq_condition}: a higher $f_{bh}^{ac}$ - or a lower coupling factor $f_{bh}^w$ - translates into a later switch of the growth mode to sub-Eddington accretion rates, yielding a longer duty cycle. Once the black hole has exited the Eddington-limited phase, an increase in $f_{bh}^{ac}$ or in the coupling factor $f_{bh}^w$ would mean less gas mass available for the subsequent time step, compensating the change in $f_{bh}^{ac}$ and $f_{bh}^w$. In this regime black hole growth is self-regulated. These parameters are in turn tuned to reproduce simultaneously both the AGN UV LF and the black hole mass function.

%#################################################################
\section{The mass assembly of early black holes and their host halos}
\label{sec_mass}
%#################################################################

In this section, we start by discussing the evolution of the black hole accretion rates (expressed in terms of the Eddington fraction) as a function of the halo mass at different redshifts in Sec. \ref{sec_bhar}. We then discuss the merger- and accretion-driven assembly of the black hole and dark matter components of early galaxies in Section \ref{sec_assembly}.

% ******************************************************************************
\subsection{Black hole accretion rates}
\label{sec_bhar}
% ******************************************************************************

\begin{figure*}
\includegraphics[width=0.9\textwidth]{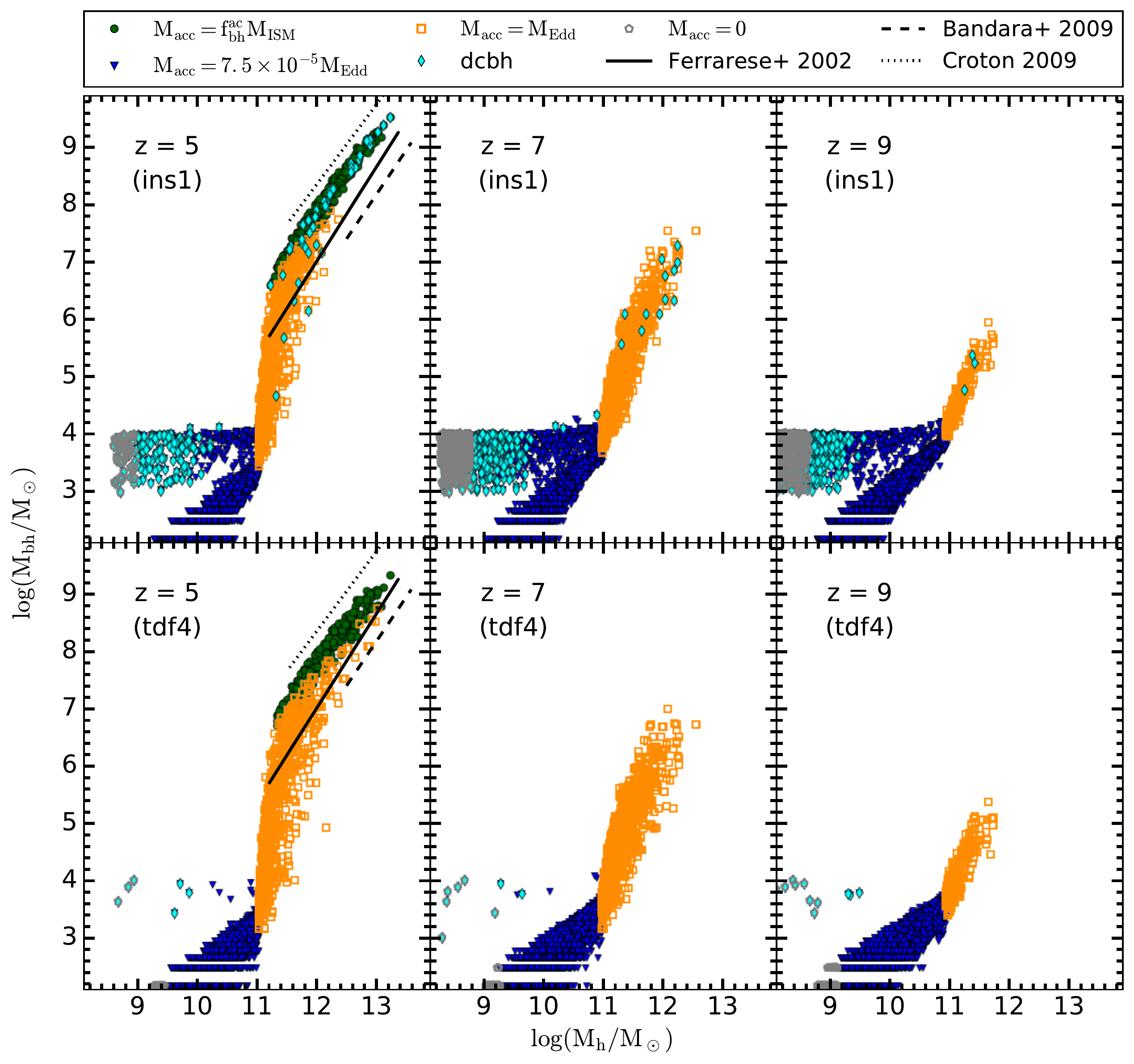}\\
\caption{The halo mass--black hole mass relation at $z = 5, 7$ and 9, as marked. The upper  and lower panels show results for the $ins1$  and $tdf4$ cases, respectively. The different symbols in each panels show the black hole accretion channels as marked in the top bar (green filled circles represent the gas-limited accretion channel, blue filled triangles are for stunted accretion mode and orange empty squares are for Eddington-limited accretion); the cyan filled diamonds represent DCBHs. In the $z=5$ panel, the solid, dashed and dotted lines show the (average) observational results inferred for the local universe \citep{ferrarese2002}, for $0.1 \lesssim z \lesssim 0.3$ \citep{bandara2009}, and theoretical predictions at $z=5$ \citep{croton2009}, respectively.}
\label{bhm_starm}
\end{figure*}

The relation between black holes and their host galaxies remains poorly constrained at high redshifts. Given that the only confirmed black holes at high redshifts are those powering luminous quasars, where the light from the accreting black hole over-shines that from the host galaxy, the stellar mass/bulge mass and the stellar velocity of the host cannot be measured. The best estimates of the dynamical mass for quasars are then obtained through measurements of the cold molecular gas properties in sub-millimetre observations \citep[e.g.][and references therein]{venemans2016, shao2017, decarli2018}. Black hole masses instead are generally inferred using rest-frame ultra-violet indicators \citep[e.g.][]{vestergaard2002, mclure2002}.  

In Fig.~\ref{bhm_starm} we show the black hole mass-halo mass relation at $z =5, 7$ and 9 for both the $ins1$ and the $tdf4$ scenarios, color-coded by the black hole accretion channel. We start by discussing the results of the $ins1$ model, which yields the upper limit of the black hole mass. Independently of redshift, black holes residing in low-mass halos ($M_h \lsim 10^9 \msun$) do not accrete at all, since SNII feedback ejects all of the gas mass from the host galaxies. The double tail at $M_h \sim 10^{9-11} \Msun$ is due to the presence of two types of black hole seeds: the upper part is comprised of (mixed) black holes resulting from the mergers of DCBHs and SBHs, while the lower part is comprised of SBHs. The black holes in this halo mass range show {\it stunted growth}, given that they accrete at a rate $\dot{M}_{bh} = 7.5 \times 10^{-5} \dot{M}_{Edd}$. Interestingly, since DCBH seeds are assigned solely on the basis of halo bias \citep[see discussion in][]{dayal2018}, they cover almost the entire halo mass range at all the redshifts studied. Once the critical halo mass threshold is reached, black holes can start accreting at the Eddington rate (\textit{Eddington-limited} phase); this phase ends when the \textit{gas-limited} condition $M_*^{gf}(z) f_{bh}^{ac} \left(1-\epsilon_r \right) < M_{Edd}(z)$ is met. This happens for the highest-mass halos with $M_h \gtrsim 10^{12} \msun$ at $z=5$. Our black hole mass-halo mass relation is therefore described by a triple power-law at $z \sim 5$: flat at $M_h \lsim M_{h}^{crit}$, almost vertical at $M_{h}^{crit} \lsim M_h \lsim 10^{11.6} \Msun$, and $\log(M_{bh}/\msun) \propto \log(M_h/\msun)^{1.4}$ at $M_h \gsim 10^{11.6} \Msun$. On the other hand, notice that black holes at $z \gsim 7$ never enter the last accretion phase. Finally, the slight increase of $M_{h}^{crit}$ with decreasing redshift naturally results in more massive halos (and hence black holes) accreting at the Eddington rate with decreasing redshift.

We then look at results from the $tdf4$ scenario that includes the impact of the UV background and a dynamical friction-induced time delay in mergers. As seen from the same figure, the $M_{bh}-M_h$ relation in this scenario follows a similar slope as the $ins1$ model for $M_h \gsim 10^{11} \Msun$ at all the redshifts considered. However, as a result of the higher LW threshold used, the total number of DCBH black hole seeds in the $tdf4$ model is lower by a factor of $\approx 50$. In addition, this model has a larger scatter (by about a factor of two) in the black hole masses at a given halo mass compared to the $ins1$ model. This is in accord with previous works \citep[e.g.][]{peng2007} that find the number of merging events to anti-correlate with the width of the scatter in the black hole mass-host mass relation. 

We find that for halos with $M_h \gsim 10^{11} \Msun$, for both models, while the slope of our $M_{bh}-M_h$ relation is consistent with a number of other (observational and theoretical) works, its normalisation is bracketed by the observational/theoretical ranges. This is quite encouraging given the different proxies used to estimate the halo masses in low-$z$ observational works, ranging from bulge properties \citep{ferrarese2002} to gravitational lens modelling \citep{bandara2009}, as well as theoretical model results extrapolating local relations to $z=5$ \citep{croton2009}.

\begin{figure*}
\includegraphics[width=0.9\textwidth]{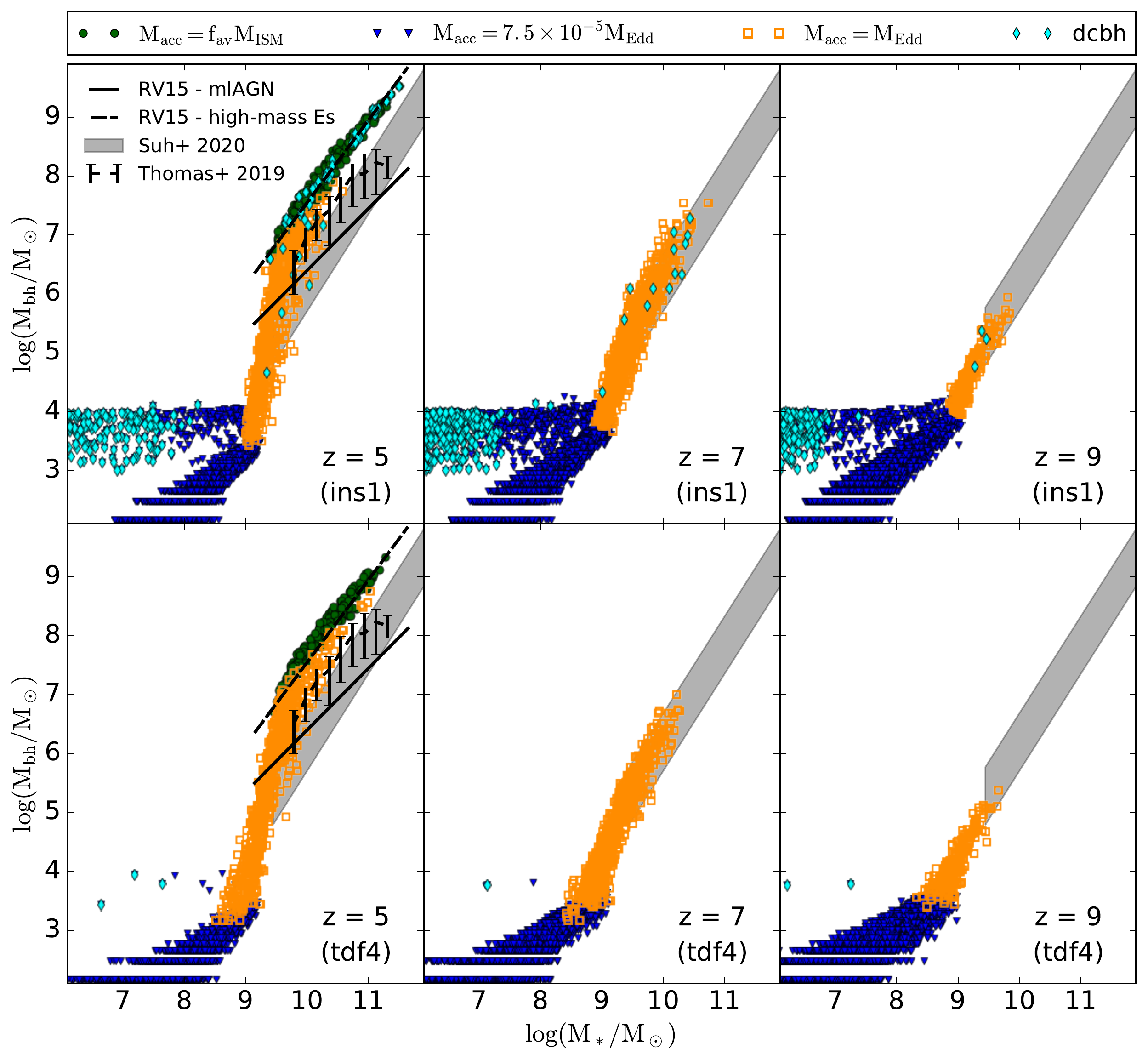}\\
\caption{The stellar mass--black hole mass relation at $z = 5, 7$ and 9, as marked. The upper and lower panels show results for the $ins1$ and $tdf4$ cases, respectively. The different symbols in each panels show the black hole accretion channels as marked in the top bar (green filled circles represent the gas-limited accretion channel, blue filled triangles are for stunted accretion mode and orange empty squares are for Eddington-limited accretion); the cyan filled diamonds represent DCBHs. In the $z=5$ panel, we plot the observational results for local moderate-luminosity AGN (solid line) and high-mass ellipticals (long-dashed line) inferred in \protect\cite{reines2015}, the results of the \textit{SIMBA} simulation at $z=5$ from \protect\cite{thomas2019} (short-dashed line), and those observed at $z=0-2$ from \protect\cite{suh2020} (grey shaded areas).}
\label{bhm_starm_app}
\end{figure*}

From an observational perspective the stellar mass is easier to measure than the halo mass, especially since telescopes like JWST will be able to estimate stellar masses also at high redshifts. Therefore, as a complement to Fig.~\ref{bhm_starm}, we show in Fig.~\ref{bhm_starm_app} the $M_*-M_{bh}$ relationship as predicted from our model. A debate has been going on in the literature regarding the evolution of this relation \citep[see the introduction of][for a review on the current state of the discussion]{suh2020}. In brief, some works inferred a positive evolution with redshift of the $M_{bh}-M_*$ ratio, suggesting that at earlier times black hole growth was more efficient than that of the host galaxies \citep[see for instance][]{peng2006, merloni2010, decarli2010, caplar2018}, while other authors found no significant difference in the mass ratios of high- and low-redshift AGN, both with observational \citep{shields2003, shankar2009, salviander2015, shen2015} and theoretical \citep{thomas2019} studies. To complicate things further, selection biases at high redshift might affect these measurements by favouring the detection of overmassive active black holes \citep{lauer2007}. From Fig.~\ref{bhm_starm_app} it appears that the $M_{bh}-M_*$ relation of the low-luminosity subset of our AGN sample, corresponding to $M_* \sim 10^{8.5-9.5} \msun$, does not evolve significantly with redshift, and is compatible with the results from \cite{suh2020}. On the other hand, galaxies with masses $M_* \gsim 10^{10} \msun$ sit $\sim 1$ dex higher with respect to the same low-redshift measurements and the relation for moderate-luminosity local AGN shown in \cite{reines2015}, pointing towards a positive evolution of the high-mass end of the $M_{bh}-M_*$ relationship. This dual behaviour of the relationship is also found in other studies \citep[e.g.][]{merloni2010}, but we point out that any comparison with observations at $z<4$ and our results cannot yield any solid conclusion, as our model does not extend to lower redshift, and any extrapolation of higher-redshift results would not take into account the different physical processes intervening in more recent epochs, such as Helium reionisation and galaxy quenching processes. For the same mass range, our relation also seems to be in very good agreement with the relation inferred for AGN in old high-mass elliptical galaxies \citep{reines2015}. In conclusion, our results point to black holes hosted by massive galaxies above $10^{9.5} \msun$ to grow efficiently and be more massive than black holes in galaxies with the same stellar mass at lower redshift, while black holes hosted in low-mass galaxies below $10^{9.5} \msun$ are less massive and follow a steeper relation \citep{volonteri2011}. The different behaviour at low and high stellar mass is simply a reflection of the behaviour at low and high halo mass.

\begin{figure}
\includegraphics[width=0.5\textwidth]{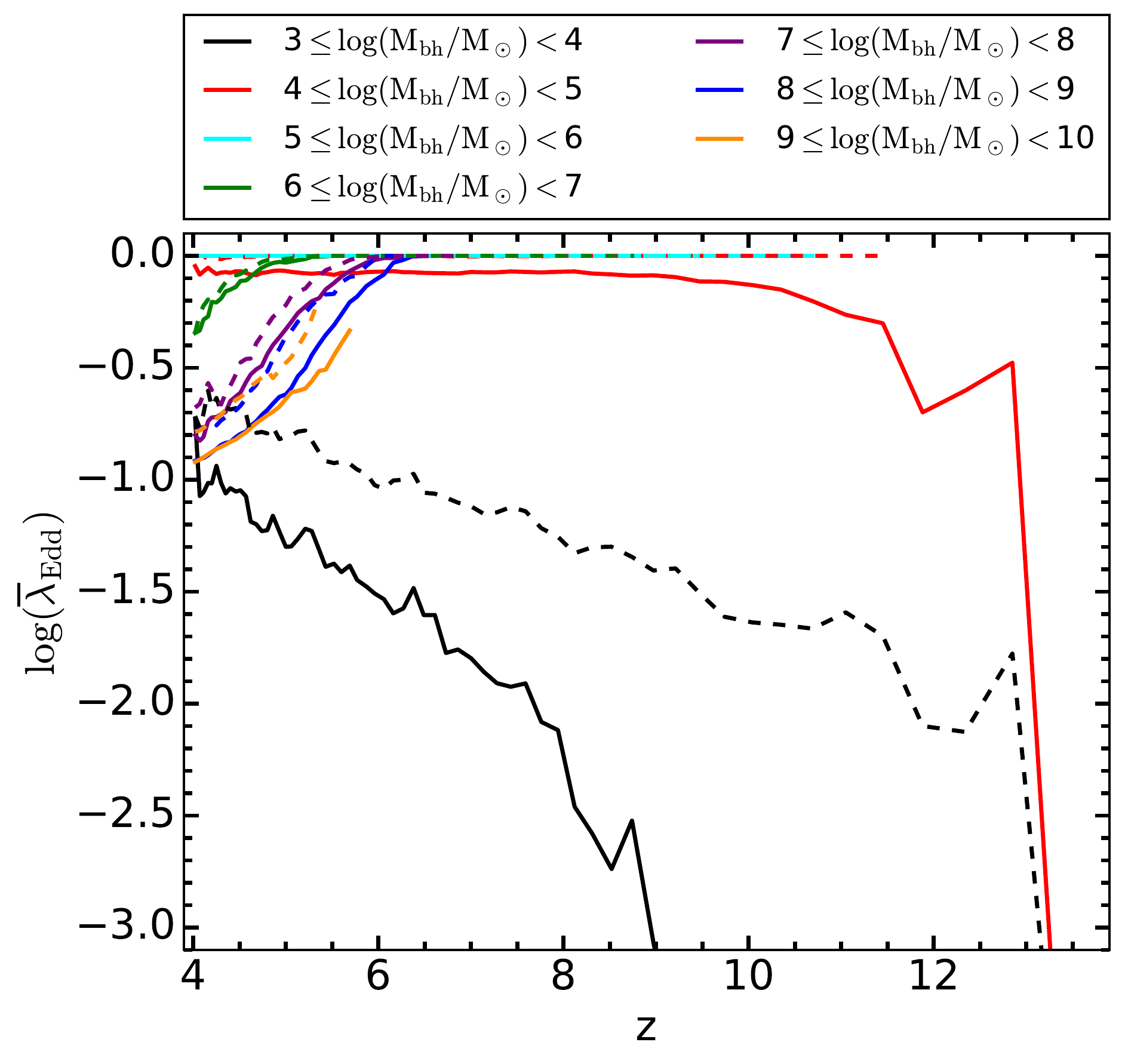}\\
\caption{The redshift evolution of the (logarithm of the) average Eddington ratio $\overline{\lambda}_{Edd}$ for the different black hole mass bins marked. Solid and dot-dashed lines show results for the $ins1$ and $tdf4$ scenarios, respectively.}
\label{average_edd_frac_pdf}
\end{figure}

With this picture in mind, we can now look at the evolution of the Eddington ratio, defined as $\lambda_{Edd} = \dot{M}_{bh}^{ac}/\dot{M}_{Edd}$, which is indicative of the black hole growth efficiency. By construction, we have that $\lambda_{Edd} = 7.5 \times 10^{-5}$ for halo masses $M_h < M_h^{crit} \sim 10^{11.5} \msun$, while $\lambda_{Edd}$ is generally between $0.1$ and 1 for black holes in higher-mass halos. The results for the average Eddington ratios ($\overline{\lambda}_{Edd}$) as a function of redshift and for different black hole mass bins are shown in Fig.~\ref{average_edd_frac_pdf}. Low-mass black holes ($M_{bh} \sim 10^{3-4} \Msun$) reside in progressively more massive halos with decreasing $z$, eventually starting to accrete at the Eddington rate (also see Fig.~\ref{bhm_starm}), leading to an increases of $\overline{\lambda}_{Edd}$ from $10^{-3}$ at $z=9$ to $0.1$ by $z=4$. Moving on, approximately $90\%$ of the black holes in the mass range $10^4-10^5 \Msun$ are in the Eddington-limited regime at $z=4$, with $\overline{\lambda}_{Edd} \approx 0.9$, while the rest of the black holes sits in low-mass halos and accretes at $\overline{\lambda}_{Edd} = 7.5 \times 10^{-5}$. Black holes with masses $10^{5-6} \Msun$ generally accrete at $\overline{\lambda}_{Edd}=1$ at all redshifts, as also seen from Fig.~\ref{bhm_starm}. As black holes grow above $10^6 \Msun$ and switch to the gas-limited accretion phase, the average Eddington ratio drops from $\overline{\lambda}_{Edd} = 1$ to progressively lower values $\sim 0.1$ \citep[see e.g.][]{merloni2004}. This is because as $M_{bh}$ increases, so does the Eddington mass, with the consequence that more massive black holes switch to the gas-limited phase at higher redshifts. Note that by construction our black hole accretion rate is capped at the Eddington rate, so we do expect a tension between our results and the observed Eddington rato distribution at high redshift, which extends to super-Eddington regimes \citep[see for instance][]{kelly2013}. In this sense, the results showed in this plot are meant to portrait a qualitative more than quantitative trend.

Compared to the $ins1$ model, in the $tdf4$ scenario, the largest difference is seen for the lowest mass bins ($M_{bh}=10^{3-5} \Msun$). We also see a slight increase in $\mathrm{\overline{\lambda}}$ for $M_{bh} \gtrsim 10^6 \msun$. This is because a delay in the baryonic mass assembly in the $tdf4$ scenario causes (i) the black holes to enter the Eddington-limited phase with a lower mass, and (ii) to stay in this phase longer, since the condition $M_{Edd}(z)<M_*^{gf}(z) f_{bh}^{ac} \left(1-\epsilon_r \right)$ is met for a longer amount of time. Additional delays in black holes mergers which are not taken into account here, such as the black hole binary inspiraling phase and the final parsec timescale, might further increase the average Eddington ratio.

%******************************************************************************
\subsection{Joint assembly of black holes and their host halos}
\label{sec_assembly}
%******************************************************************************

\begin{figure}
\includegraphics[width=0.48\textwidth]{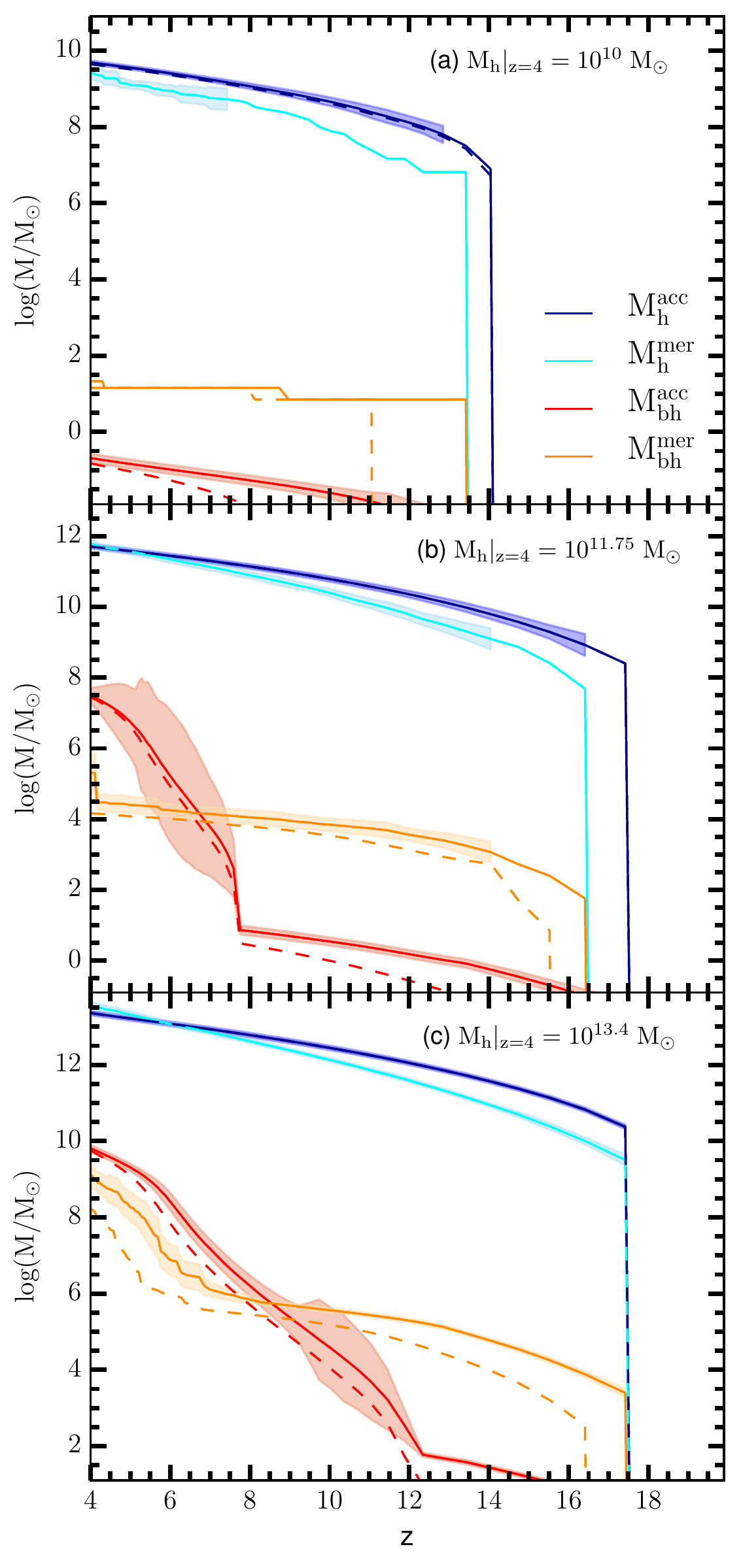}\\
\caption{The cumulative mass assembly of the black hole and the host dark matter halo as a result of mergers ($M_{bh}^{mer}$ and $M_h^{mer}$, represented respectively by the orange and the cyan lines) and accretion ($M_{bh}^{acc}$ and $M_h^{acc}$, red and blue lines) as a function of redshift. We show results for $z=4$ halo masses in the range $M_h \sim 10^{9.9-10.1} \Msun$ (upper panel), $M_h \sim 10^{11.65-11.85} \Msun$ (middle panel) and $M_h \sim 10^{13.3-13.5} \Msun$ (lower panel), as marked, averaging over $21$ halos in each panel. The solid and dashed lines in each panel show results for the $ins1$ and $tdf4$ cases, respectively; shaded regions show the 1-$\sigma$ variance for the $ins1$ scenario. As expected, while the halo assembly is identical in both models, black hole assembly is delayed in the $tdf4$ model.}
\vspace{-1mm}
\label{mah_bh_h}
\end{figure}

\begin{figure}
\includegraphics[width=0.48\textwidth]{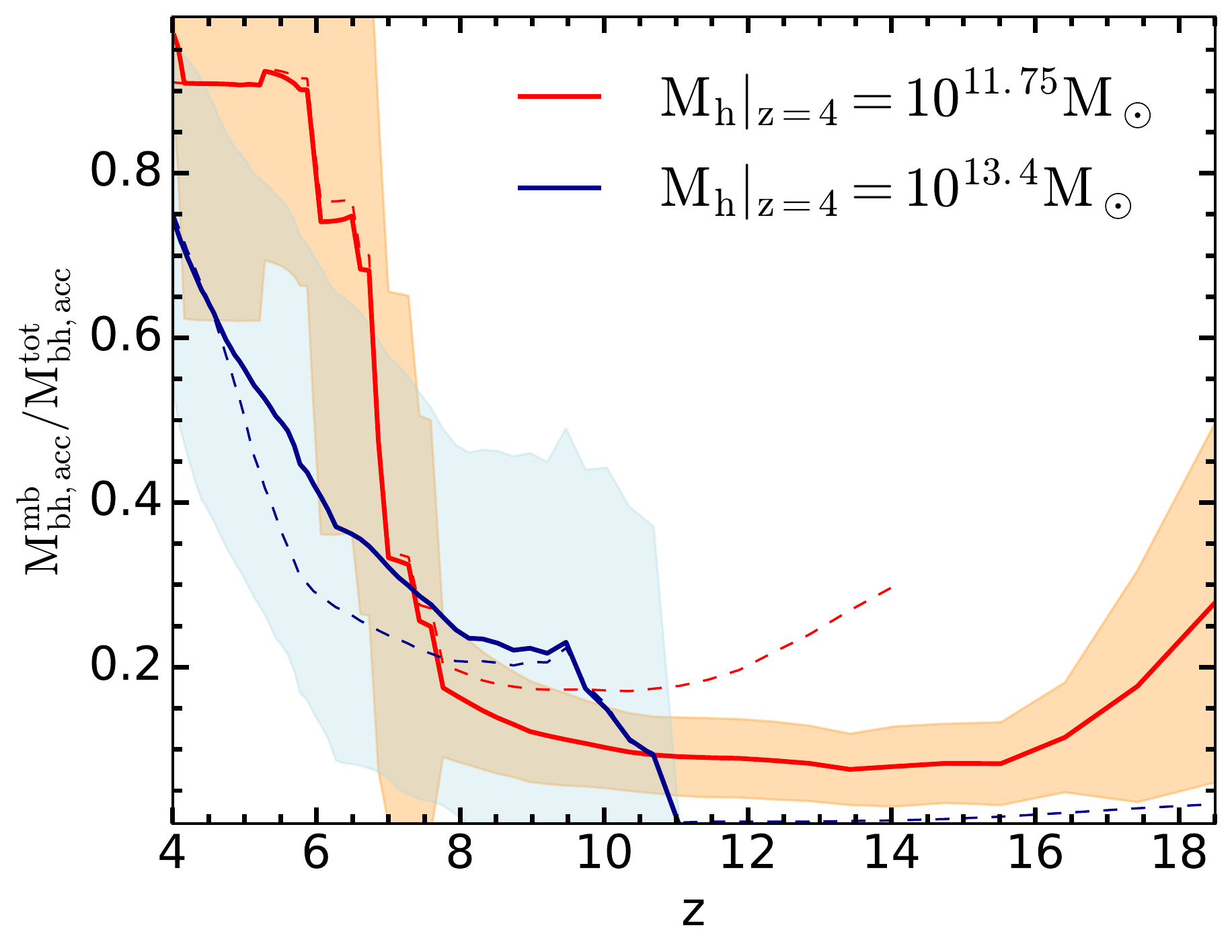}\\
\caption{The redshift evolution of the ratio of the cumulative mass accreted by black holes in the major branch with respect to that accreted in the whole merger tree of a galaxy (averaged over $21$ halos) for $z=4$ halo masses of $M_h \sim 10^{11.65-11.85} \Msun$ (red lines) and $M_h \sim 10^{13.3-13.5} \Msun$ (blue lines), as marked. The solid and dashed lines in each panel show results for the $ins1$ and $tdf4$ cases, respectively; shaded regions show the $1-\sigma$ variance for the $ins1$ scenario.}
\label{bh_mb}
\end{figure}

We now study the relative contribution of mergers and accretion to the joint assembly of black holes and their host halos at $z>4$. For both black holes and halos, the merged mass is that brought in by all the progenitors of the previous $z$-step; the accreted mass is the ISM and IGM mass that is accreted onto the black hole and halo, respectively. Here we explore results for 21 galaxies over three halo mass bins (at $z=4$) each: low-mass halos ($M_h \sim 10^{9.9-10.1} \Msun$), intermediate mass halos ($M_h \sim 10^{11.65-11.85} \Msun$) and high-mass halos ($M_h \sim 10^{13.3-13.5} \Msun$). We start by discussing results from the $ins1$ scenario. Naturally, both the $ins1$ and $tdf4$ scenarios show the same results for halo assembly; the dynamical friction delay only affects the baryonic component.

Low-mass halos (panel a of Fig.~\ref{mah_bh_h}) start assembling their dark matter mass at $z \sim 14$. Most of the halo growth is driven by accretion down to $z \sim 4$, at which point mergers start contributing to the halo mass assembly in equal measure. On the other hand, their black hole assembly is dominated by stochastic mergers across the whole redshift range $z \sim 13-4$. The accretion mode is sub-dominant given the negligible accretion rates ($\sim 7.5\times 10^{-5} M_{edd}$), and contributes less than $10\%$ of the total black hole mass by $z \sim 4$.

Intermediate-mass halos (panel b of the same figure) naturally start assembling earlier, at $z \sim 17.5$. Accretion dominates the mass assembly until $z \sim 7$, below which point mergers start contributing equally. While mergers dominate the black hole mass build-up down to $z \sim 7$, thereafter Eddington-limited accretion takes over, dominating the black hole mass budget by $\sim 2$ orders of magnitude by $z=4$. By then, more than $95\%$ of this black hole accretion has taken place in the major branch (see Fig.~\ref{bh_mb}), wherein black holes accrete in the Eddington-limited phase. The high variance here is due to the fact that different halos, with their different assembly histories, overcome the critical halo mass for Eddington-limited accretion at different redshifts.

\begin{figure}
\includegraphics[width=0.48\textwidth]{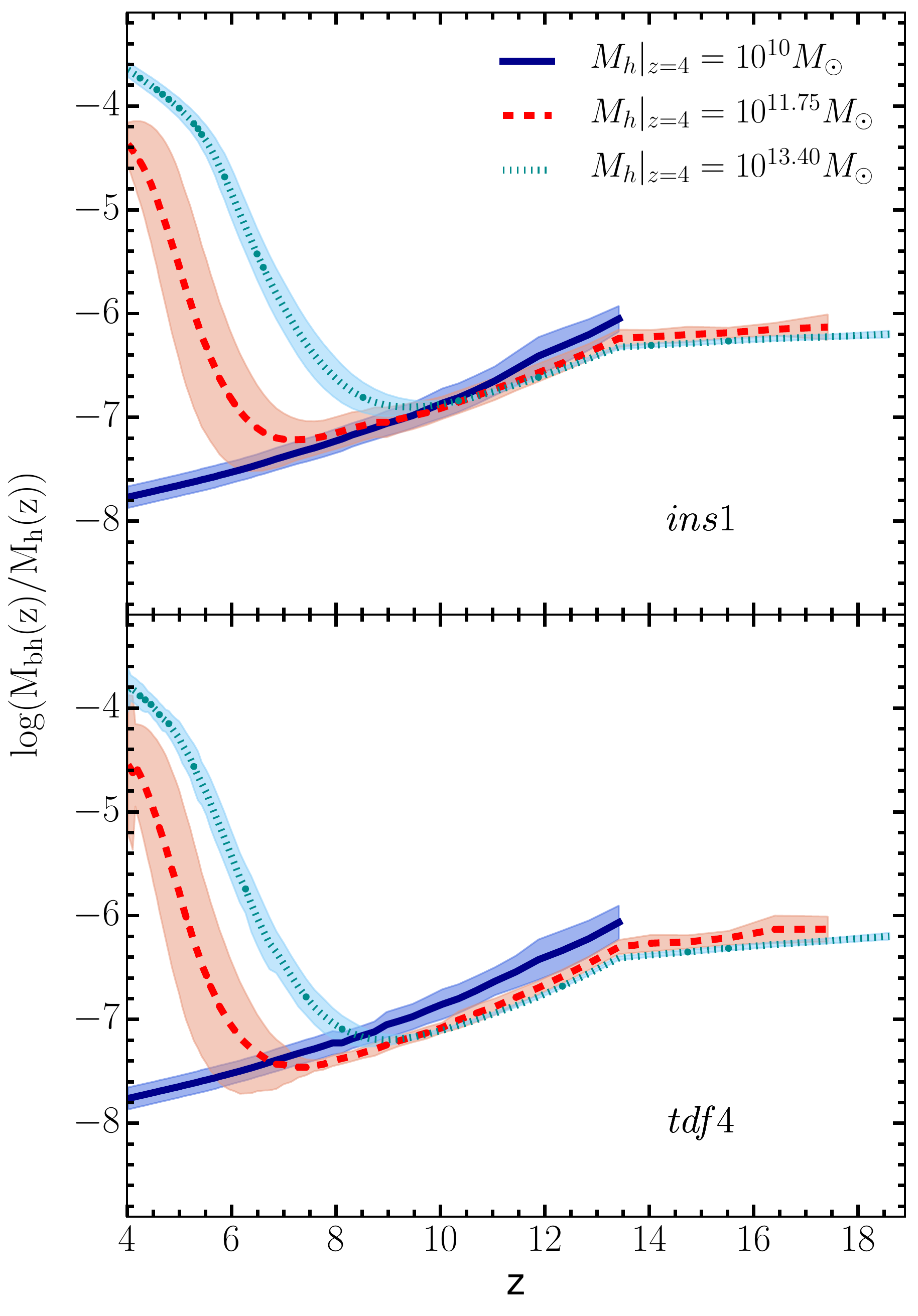}\\
\caption{As a function of $z$, we show the ratio of the cumulative black hole mass to the cumulative halo mass (averaged over $21$ halos) for $z=4$ halo masses of $M_h \sim 10^{9.6-10.1} \Msun$ (solid dark blue line), $M_h \sim 10^{11.65-11.85} \Msun$ (dashed red line) and $M_h \sim 10^{13.3-13.5} \Msun$ (dotted light blue line), as marked. The upper and lower panels show results for the $ins1$ and $tdf4$ cases, respectively; shaded regions show the associated $1-\sigma$ variance.}
\label{mbh_over_mh}
\end{figure}

The dark matter mass assembly of high-mass halos is dominated by accretion until $z \sim 6$, below which mergers take over as the main growth mechanism \citep[see also][]{cattaneo2002}. As for black holes, mergers dominate (and increase the mass by about three orders of magnitude) between $z \sim 17.5-9$; accretion dominates instead the later phases, contributing about 90\% of the final mass at $z \sim 4$. It is interesting also to notice that for high-mass halos the variance in the merged components is generally lower, and significant only at $z \leq 7$, when stochastic major mergers take place. Further, more than one progenitor is able to overcome $M_{h}^{crit}$, and on average only about $20\, (75\%)$ of the mass in assembled in the major branch at $z \sim 10 \, (4)$ as shown in Fig.~\ref{bh_mb}.  

In the $tdf4$ case the black hole mass assembly proceeds slower, due to fewer mergers (because of dynamical friction) that bring in lower gas masses (because of the photo-evaporation of gas from low-mass halos).

To summarise, while black holes residing in low-mass halos can grow only through (rare) mergers, those in intermediate- and high-mass mass halos predominantly grow through mergers only in the early phases, before switching to an accretion-dominated growth phase. As expected, mergers are less important in the $tdf4$ scenario, contributing $\sim 10^{-3.5}\%~(10^{-2}\%)$ of the total mass in intermediate (high) mass halos.

Finally, in Fig.~\ref{mbh_over_mh} we plot the redshift evolution of the ratio of the cumulative black hole mass to the cumulative halo mass for the same halo mass bins as in Fig.~\ref{mah_bh_h}. As shown, black holes in low-mass halos (when present) are starved throughout the redshift range $z = 14-4$, driving the mass ratio to a constant decrease. Intermediate-mass halos show a flat evolution at $z>13$, when the black hole seed formation efficiency is still $100\%$ (i.e.\ every newly-born halo has a black hole). At $z \sim 13 -6$, the ratio drops as new halos form without black hole seeds. At $z\lsim 6$, the ratio shows a steep rise as the black holes residing in the biggest progenitors enter the Eddington-limited accretion phase. This results in a rise from $M_{bh}/M_h \sim 10^{-7.4}$ at $z=6$ to $M_{bh}/M_h \sim 10^{-4.5}$ at $z=4$. While the evolution of the highest mass halos is qualitatively similar to intermediate mass halos, their black holes already enter the Eddington-limited growth channel at $z \approx 8$. Sustaining this growth for a longer time, they show a mass ratio that increases from $M_{bh}/M_h \sim 10^{-6.8}$ at $z=8$ to $M_{bh}/M_h \sim 10^{-3.6}$ at $z=4$. Again, intermediate mass halos show a larger variance with respect to high-mass halos as a result of the diverse merging and accretion histories of their black holes.

%#################################################################
\section{Key properties of the black hole population in early galaxies}
\label{sec_res}
%#################################################################

We now explore the key properties of BHs in early galaxies, specifically focusing on their occupation fraction and duty cycles in this section. 

\begin{figure}
\includegraphics[width=0.5\textwidth]{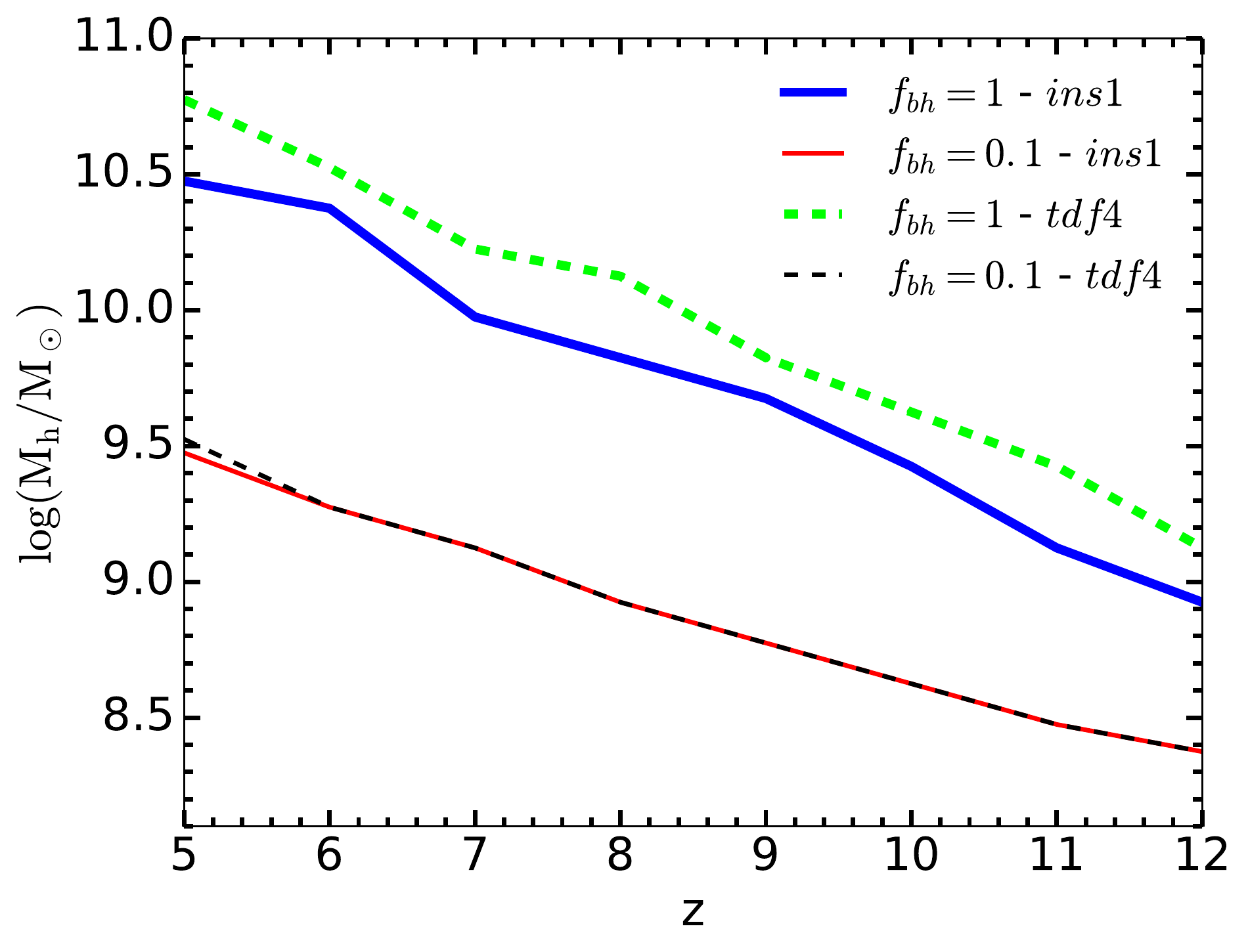}\\
\caption{The redshift evolution of the minimum halo mass for which the central black hole occupation fraction $\mathrm{f_{bh}} = 1$ (thick lines) and $\mathrm{f_{bh}} = 0.1$ (thin lines) for both the $ins1$ (solid lines) and the $tdf4$ (dashed lines) scenarios, as marked.}
\label{fbh_unity}
\end{figure}

\begin{figure*}
\includegraphics[width=0.9\textwidth]{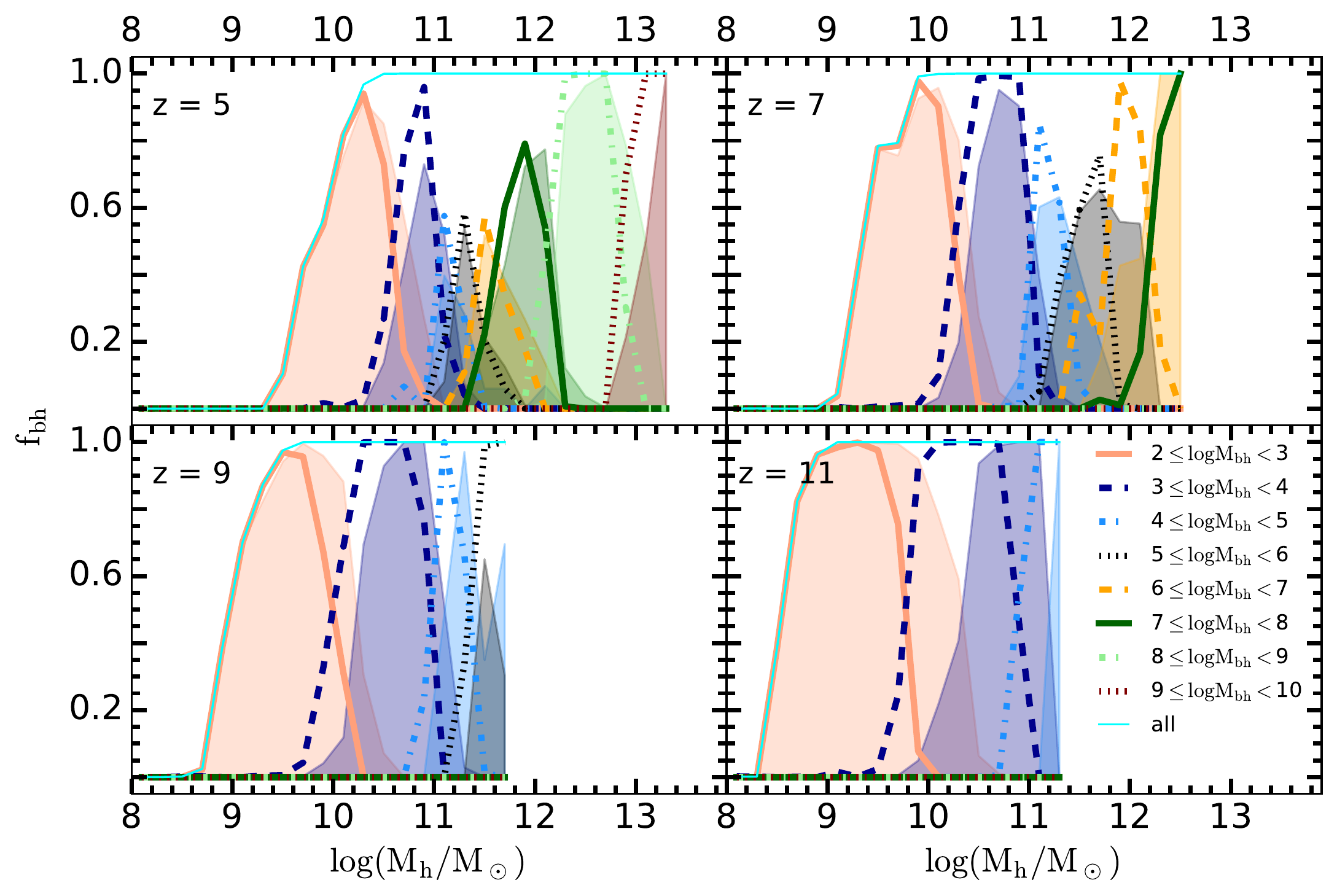}\\
\caption{The central black hole occupation fraction ($\mathrm{f_{bh}}$) as a function of halo mass for the different black hole mass bins noted. The occupation fractions are shown using lines for the \textit{ins1} case and as shaded areas for the \textit{tdf4} case. The solid thin cyan line shows the cumulative black hole fraction for the $ins1$ case.}
\label{bh_occfrac}
\end{figure*}

% ******************************************************************************
\subsection{The black hole occupation fraction}
\label{sec_bhoccfrac}
% ******************************************************************************

We define the black hole occupation fraction ($\mathrm{f_{bh}}$) as the fraction of halos hosting a {\it central} black hole. In the case of merging systems, this refers to the fraction of central galaxies hosting a black hole. We note that this specification is important only in the $tdf4$ scenario; in the $ins1$ case galaxies merge at the same time as their halos. In our model, halos are populated with SBH seeds down to $z=13$. Lower-$z$ halos contain a black hole if they are either seeded with a DCBH or if they gain one through mergers. The latter (and dominant) mechanism naturally implies that the minimum halo mass for which $\mathrm{f_{bh}} = 1$ increases with decreasing redshift. Indeed, as shown in Fig.~\ref{fbh_unity}, we find that $\mathrm{f_{bh}}=0.1$ (i.e. 10\% of halos contain a central black hole) for a minimum halo mass of $M_h \sim 10^{8.4} \Msun$ at $z \sim 12$, which increases to $M_h \sim 10^{9.5} \Msun$ by $z=5$. Reaching $\mathrm{f_{bh}}=1$ naturally requires a higher halo mass of $M_h \sim 10^{8.9}\, (10^{10.5}) \Msun$ at $z\sim 12\, (5)$. As seen from the same plot, there is no sensible difference in the $ins1$ and $tdf4$ results for $\mathrm{f_{bh}}=0.1$. However, the minimum halo mass for $\mathrm{f_{bh}}=1$ is about $0.3-0.4$ dex higher in the $tdf4$ model as compared to $ins1$ model. This is because a merging galaxy and its black hole reach the central galaxy at a later time step as compared to the instantaneous ($ins1$) scenario.

\begin{figure*}
\includegraphics[width=0.9\textwidth]{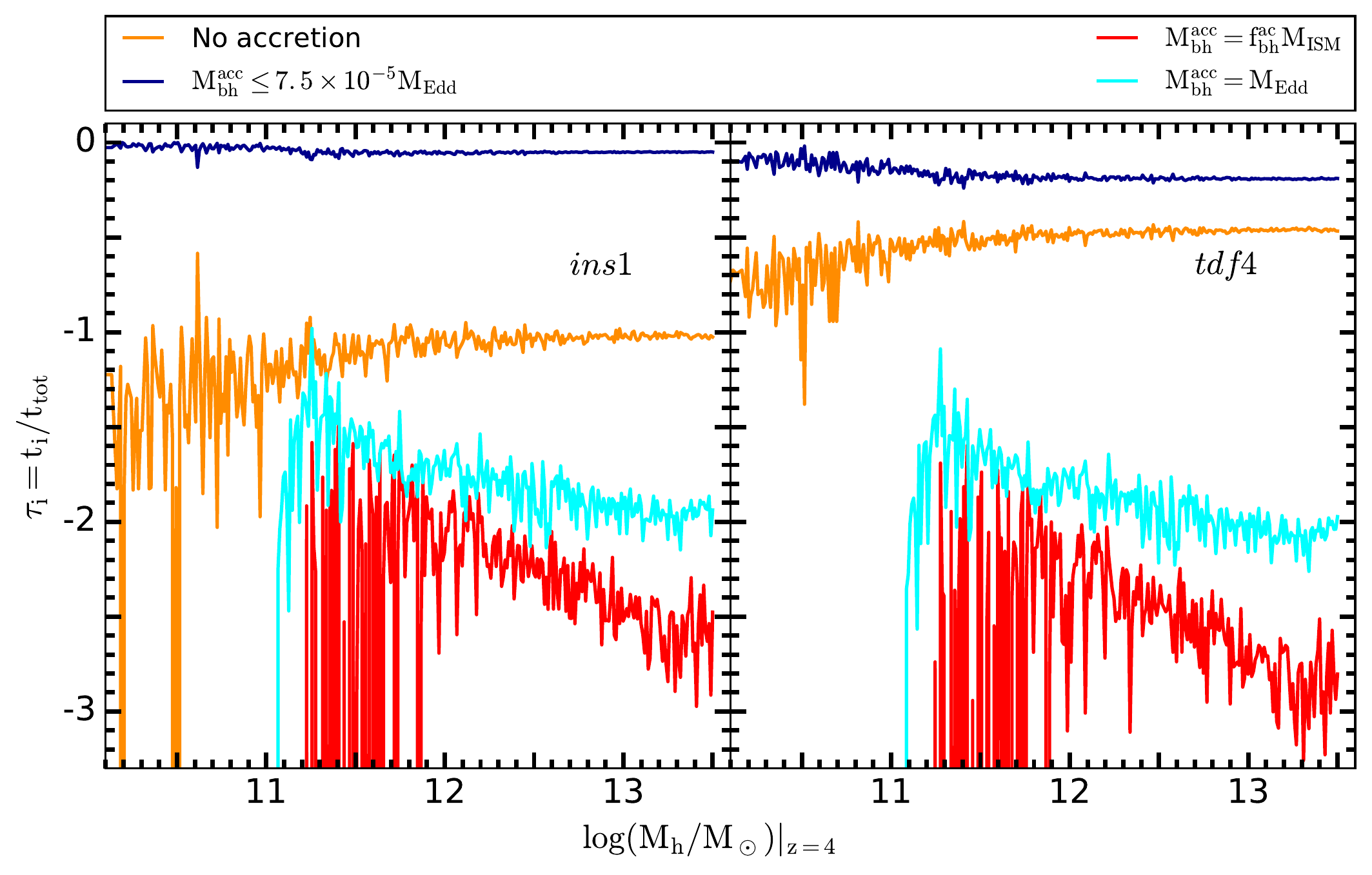}\\
\caption{Logarithm of the fraction ($\tau_i$) of the cumulative lifetime of the black hole spent in the stunted (blue line), Eddington-limited (cyan line) and gas-limited (red line) accretion modes as a function of the $z=4$ halo mass ($M_h|_{z=4}$). The orange line shows black holes in SNII-feedback dominated galaxies that do not accrete at all.}
\label{bh_tweighed_dutycycle}
\end{figure*}

We then now deconstruct $\mathrm{f_{bh}}$ as a function of halo mass for different black hole mass bins, and show the results in Fig.~\ref{bh_occfrac}. Firstly, as expected, independent of the redshift and model used, there is a clear trend of black holes of increasing mass being hosted in increasingly massive halos. In terms of the most massive black holes, while by $z \simeq 9$ the most massive halos ($M_h \sim 10^{11.5} \msun$) host black holes as massive as $\sim 10^6 \Msun$, this increases to $M_{bh} \sim 10^{9.5} \Msun$ by $z=5$ for halos with $M_h \sim 10^{13.4} \msun$. At each redshift the overlap between the different occupation fraction curves is indicative of the intrinsic scatter of the $M_{bh}-M_h$ relation shown in Fig.~\ref{bhm_starm}. Indeed, while at $z = 9$ halos with $M_h \sim 10^{11-12} \Msun$ host black holes with $M_{bh} \sim 10^4-10^6 \Msun$, at $z=5$ the same halo masses host black holes spanning four orders of magnitude in mass, between $10^{3.5}-10^{7.5} \Msun$. This is mainly due to the black holes residing in these halos entering the Eddington-limited accretion regime. Further, deviations from the average $M_{bh}-M_h$ relation, which depend on the black hole mass at the time the halo reaches $M_{h}^{crit}$, are amplified by the exponential growth of black holes during the Eddington-limited growth phase.

In the $tdf4$ case, halos of a given mass range typically host black holes of slightly lower masses compared to $ins1$. This is due to a combination of delayed mergers and (to a lower extent) lower gas masses, given that mergers are the dominant mode of black hole growth at $z \gsim 9$ in intermediate- and high-mass halos. As accretion overtakes the mass build-up at lower $z$, results from the $ins1$ and $tdf4$ models come into closer agreement.

% ******************************************************************************
\subsection{Time-weighed duty cycle}
\label{sec_duty_cycle}

% ******************************************************************************

We now explore the black hole duty cycle defined as the fractional lifetime a black hole spent in the stunted, Eddington-limited and gas-limited accretion modes. This is done for all the $z=4$ halos that host a black hole, accounting for all of the progenitors along the merger tree. The time-weighed duty cycle for each accretion mode is then defined as $\tau_i = t_i/t_{tot}$, where $t_i$ indicates the time spent in each of the three accretion channels ($i$) and $t_{tot}$ is the total lifetime of the black hole. 

As expected in a hierarchical growth scenario, most of the lifetime of the progenitors of the final black hole is spent in small halos, where they are subject to stunted accretion. In particular, black hole progenitors of halos with $M_h|_{z=4} < M_{h}^{crit}$ spend all of their lifetime in a stunted accretion mode or -- if they are in SN-feedback dominated halos (which are more common among the progenitors of high-mass halos) -- not accreting at all. For $M_h|_{z=4} \gtrsim 10^{11.5} \msun$ halos, the time spent in both the Eddington and gas-limited regimes decreases with increasing halo mass, again due to an increasing fraction of low-mass halos in their merger trees. The black holes in intermediate-mass halos spend at most 5\% and 1\% of their lifetime accreting in the Eddington- and gas-limited modes, respectively; these drop to 0.1\% and 1\% for the highest mass halos.

The most important difference upon implementing dynamical friction delay and UV feedback is that the black hole lifetime spent in non-accreting halos increases by at least a factor of 3 for the highest-mass halos (and up to an order of magnitude for the lowest-mass halos) due to UV suppression of their gas mass.

Finally, the variations (up to one order of magnitude) between the duty cycles of halos of very similar final masses is driven by the assembly histories of both black holes and their host halos, which determine both the halo and black hole masses, as well as the amount of gas mass available for accretion onto the black hole. 

% ******************************************************************************
\subsection{Black hole predictions at $z \geq 5$}
\label{bhmf}
% ******************************************************************************

\begin{figure*}
\includegraphics[width=\textwidth]{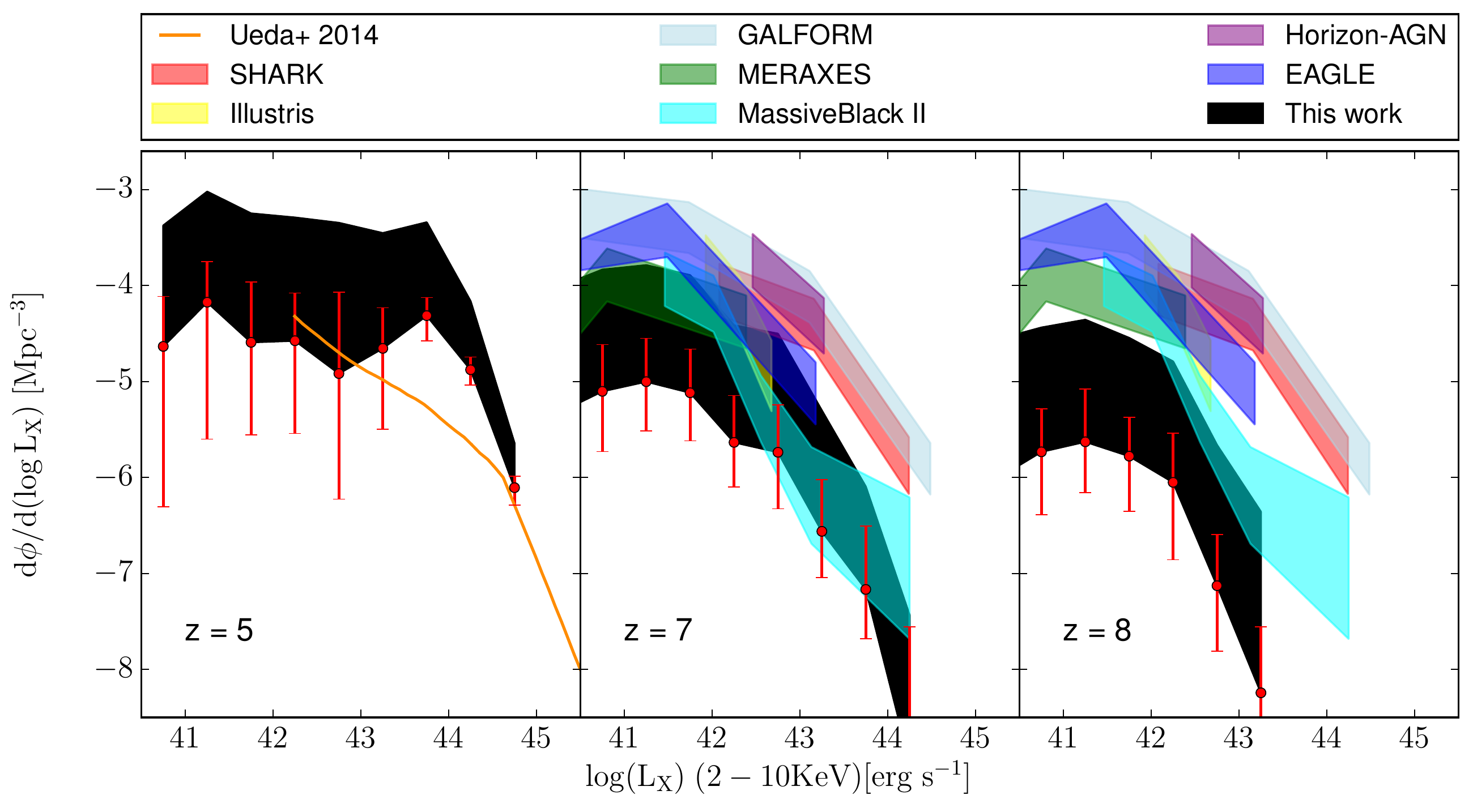}\\
\caption{Evolution of the AGN X-ray luminosity function at $z = 5, 7, 8$. The black shaded areas represent our results, for which the upper and lower limit are set respectively by the intrinsic and the obscured XLF, where the obscured fraction as a function of luminosity is taken from \protect\cite{ueda2014}. The red error bars represent the $1-\sigma$ uncertainties on our lower limit. The other coloured shaded areas represent results from various models taken from Fig.~3 of \protect\cite{amarantidis2019}, their upper and lower limits corresponding to the minimum and maximum obscuration effect. The orange solid line shows the observational results from \protect\cite{ueda2014}.}
\label{xlf}
\end{figure*}

\begin{figure*}
\includegraphics[width=\textwidth]{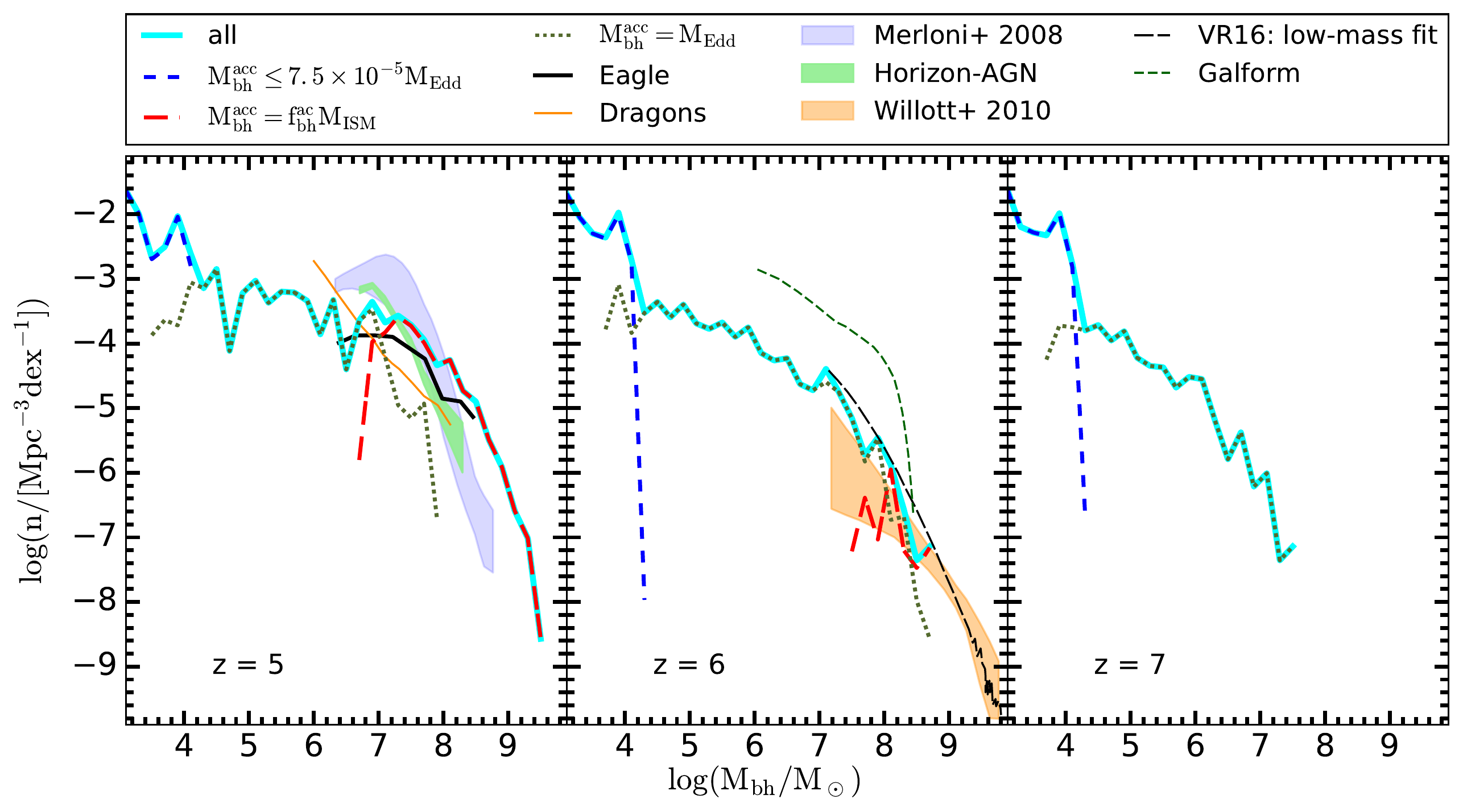}\\
\caption{Redshift evolution of the black hole mass function for the $ins1$ scenario (solid cyan line). This is deconstructed to show the mass function for black holes accreting in the stunted mode (dashed blue line), Eddington-limited mode (long-dashed red line) and the gas-limited modes (green dotted line). We also show results from other theoretical and observational works at: {\it $z=5$} from the Eagle simulations \protect\citep[][black solid line]{rosasguevara2016}, the \textit{DRAGONS} model \protect\citep{qin2017}, the Horizon-AGN simulation \protect\citep[][green shaded area]{volonteri2016b} and \protect\cite[][purple shaded area]{merloni2008}, and at {\it $z=6$} we show theoretical results from the \textit{GALFORM} model \protect\citep[][green dashed line]{griffin2019} and from \citet[][black dashed line]{volonteri2016a} and observational results \citep[][orange shaded area]{willott2010}.}
\label{bhmf_decomp}
\end{figure*}

We conclude our analysis by proposing a couple of testable predictions of our model. In Fig.~\ref{xlf} we show the resulting X-ray luminosity function (XLF) at $z = 5, 7$ and 8. Since the model provides us with the bolometric luminosity information of each AGN, we build the luminosity function just by binning our AGN sample. We adopt the \cite{duras2020} X-ray bolometric correction, while the obscured fraction as a function of luminosity is taken from \cite{ueda2014}. We compare our results to the $z=5$ observational XLF from \cite{ueda2014} (left panel), and we also compare the $7 < z < 8$ results of other SAMs and simulations taken from Fig.~3 of \cite{amarantidis2019} to our XLF at z = 7 (central panel) and z = 8 (right panel). We use our intrinsic and obscured X-ray luminosity functions as upper and lower limit on the XLF, since for the other results considered here the upper and lower limits correspond to the estimations for the maximum and minimum obscuration effects. We point out that our lower limit represents the median obscured XLF obtained from 1000 realisations of the obscuration mechanism: for each realisation, in each luminosity bin, we randomly draw a subsample of the AGN populating that bin corresponding to the obscured fraction and manually set their AGN luminosity to zero. The red error bars represent the 16th and 84th percentiles of the resulting set of obscured XLFs. We can easily notice in the z=7 and z=8 panels that introducing the threshold halo mass $M_h^{crit}$ for black hole growth delays the build-up of our XLF with respect to that of the other models, but this delay disappears at z=5 when comparing it to the observational results.

Together with the luminosity function, the mass function is the other main predictable statistical observable, though as of now it is still very loosely constrained at high redshift \citep{willott2010}. In Fig.~\ref{bhmf_decomp} we show our $z=5-7$ black hole mass function. Here we focus on the results from the $ins1$ model and deconstruct it to show the contributions from all three black hole accretion channels. The general behaviour follows that expected from the hierarchical structure formation model: black holes with $M_{bh} \lsim 10^7 \msun$ follow a power law with their number densities being virtually constant at $z \sim 5-7$. On the other hand, for $M_{bh} \gtrsim 10^7 \msun$, the number density of high-mass black holes show an exponential cut-off, specially at $z \sim 5-6$, in addition to showing a rapid evolution between $z \sim 6-7$. In particular, the mass bin $10^6 \Msun \lesssim M_{bh} \lesssim 10^7 \Msun$ experiences the greatest variation of number density in the shortest timescale, increasing by almost three orders of magnitude between $z=7$ and $z=6$ (i.e.\ in less than $300 \myr$). As a consequence of the different evolution in number densities of the low- and high-mass ends of the BHMF, at $z=5$ black holes with $M_{bh} \sim 10^{4.5} \Msun$ and $M_{bh} \sim 10^7 \Msun$ have very similar number densities. The exponential growth of black holes in the Eddington-limited accretion regime (i.e.\ all black holes with $M_{bh} \gtrsim 10^4 \msun$) makes them move from one black hole mass bin to the next one in a very short time span, while the halo mass grows much slower. Therefore, the shallow slope of the BHMF in the intermediate mass range $10^4 \msun < M_{bh} < 10^7 \msun$ at $z=5$ is a direct effect of the critical halo mass $M_{h}^{crit}$ imposed in our model.

As noted before, we reiterate that the black hole accretion channel strongly depends on its mass: while low-mass black holes with $M_{bh} \sim 10^4 \msun$ accrete at very low ($7.5 \times 10^{-5}$) Eddington rates, intermediate-mass black holes can accrete at the Eddington rate while the most massive black holes show a variable Eddington ratio.

In the same figure, we also compare our BHMF to other theoretical and observational works. At $z=6$, our results are in qualitative agreement with the observational BHMF built (using spectroscopically-selected quasars) by \cite{willott2010}, as well as the theoretical results of \cite{volonteri2016a}, based on extrapolating the observed $z=0$ black hole--host galaxy correlations to higher redshifts. At $z=5$, our results are in reasonable agreement with those from the \textit{EAGLE} simulation \citep{rosasguevara2016}. We are also in qualitative agreement with the results from the \textit{Horizon-AGN} project \citep{volonteri2016b} and the AGN synthesis model by \cite{merloni2008}, although these models seem to hint to a characteristic knee mass lower by $\approx 0.5$ dex. Finally at $z=5-6$, our results are within the range bracketed by the \textit{GALFORM} \citep{griffin2019} semi-analytic model (higher than our BHMF by more than 1 dex at $z=6$ for $10^6 \msun \lsim M_{bh} \lsim 10^8 \msun$) and the \textit{Dragons} semi-analytic model \citep{qin2017}, which predicts a lower number density (by about 0.5 dex for $10^6 \msun \lsim M_{bh} \lsim 10^8 \msun$ at $z=5$). This mismatch is indicative of the different black hole seeding and growth mechanisms implementations. \cite{griffin2019} assume that black hole growth proceeds proportionally to the star formation rate in the host galaxy, and that accretion episodes can occur also during galaxy mergers able to drive gas towards the central regions. Given the absence of any halo mass threshold for the growth of black holes, they start assembling at earlier times with respect to our model. We then expect their BHMF to evolve faster than ours at higher redshift. In \cite{qin2017}, instead, black holes are allowed to accrete a (variable) fraction of the hot gas present in the galaxy, while cold gas can only be accreted only during mergers; there is no such distinction between cold and hot gas accretion in our model, which might explain why we produce more massive black holes.

The fact that we are able to reproduce very well the observed black hole mass function at $z=5$, but at the same time finding a slight mismatch between our X-ray luminosity function and the one inferred at $z=5$ by \cite{ueda2014} at a luminosity of $L_X \sim 10^{44}\ \mathrm{erg\ s^{-1}}$, might point to a tension between these results. Nevertheless, the different AGN samples, selection criteria and corrections used in \cite{ueda2014} and \cite{willott2010} to compute respectively the X-ray luminosity function and the black hole mass function do not allow us to draw any solid conclusion about this. 

%#################################################################  
\section{Conclusions and discussion}
\label{conclusions}
%#################################################################

In this work, we have used the \textit{Delphi} semi-analytic model to study the joint assembly of high-redshift ($z>4$) black holes and their host halos. We implement a physically-motivated black hole growth mechanism that includes stunted growth in (gas-poor) low-mass halos while high-mass halos above the critical halo mass $M_{h}^{crit}(z)$ can accrete at the Eddington rate. After tuning the parameters of the model for the fiducial $ins1$ scenario (see Table~\ref{table_models}), we also consider a more realistic second case ($tdf4$) in which the mergers of the baryonic components are delayed by a dynamical friction timescale, a maximal UV feedback scenario is implemented and the LW BG threshold for DCBH seed formation is higher. We summarise here the key results.

\begin{enumerate}
\item First of all, as a consequence of our seeding mechanism and accretion model, in halos with $M_h > M_h^{crit}$ the black hole mass is assembled mostly via mergers until $z \sim 8$. Notably, this remains true even when the delay in mergers due to dynamical friction is added to the model. In fact, at such high redshift the mass ratios between halos are on average closer to unity, and the merging timescales are shorter. At $z<8$ accretion takes over. 
\item By $z=4$ we find that in halos with $M_h \approx 10^{11.75}~(10^{13.4}) \msun$ up to 95\% (75\%) of the final black hole mass is accreted within the major branch of the merger tree, while the rest comes from mergers with black holes hosted in secondary branches. While bigger halos have more progenitors growing above $M_h^{crit}$, in intermediate-mass halos only the major branch of the tree is able to fuel Eddington-limited accretion. 
\item Finally, the average Eddington ratio $\overline{\lambda}_{Edd}$ seems to have a different evolution for black holes of different masses: for $M_{bh} \gtrsim 10^5 \msun$ we have $\overline{\lambda}_{Edd} \approx 1$ at $z > 6$, while it decreases down to $\approx 0.1$ at lower redshift. Low-mass black holes with $M_{bh} < 10^5 \msun$, instead, show an opposite trend, increasing from $\overline{\lambda}_{Edd} << 1$ at high redshift to $\overline{\lambda}_{Edd} \approx 0.1-1$ at $z = 4$. 
\end{enumerate}

We tested our black hole seeding mechanism, finding that varying the initial DCBH seed mass our results remain effectively unchanged. This is in agreement with previous works \citep[e.g.][]{ricarte2018} that conclude that a hybrid scenario with a bimodal black hole seed mass distribution (such as ours) yields basically the same results as a light-seed one, given the low number density of DCBHs compared to SBHs.

To conclude, the reader should keep in mind a few caveats. First, we are not accounting for gravitational recoil kicks in mergers between massive black holes, which could either offset the central (merged) black hole or eject it altogether, hampering its subsequent growth \citep{blecha2016}. Nevertheless, spin alignment between the merging black holes might reduce the effectiveness of this mechanism, especially if the merging system is embedded in a circumbinary gaseous disk \citep{dotti2010}. We also point out that the black hole merging timescales are still probably severely underestimated: not only we do not model the black hole binary inspiralling phase after the galaxies have merged, but we are also not accounting for the final parsec problem, which requires stellar scattering processes for allowing the black holes to coalesce together. In addition, we do not consider that in this low-mass regime the stellar and gaseous components can actually drive the dynamics of the black holes, introducing stochasticity in their orbits and possibly affecting the black hole merger rate, as shown by \cite{pfister2019}. This last mechanism might widely stretch or even prevent black holes from merging at all. We should also mention that the intensity of the Lyman-Werner background necessary to form DCBH seeds is still very poorly constrained in the literature, and values of $\alpha$ higher (lower) than the ones we have chosen in Table~\ref{table_models} would result in a lower (higher) number density of DCBH seeds forming at $8 < z < 13$. Finally, we manage to form black holes as massive as $10^{10} \msun$ at $z=5$ by allowing them to accrete a fraction of the total gas mass present in the host galaxy, rather than only the gas confined in the central regions. For these reasons the black hole masses that our model $ins1$ yields should be considered as an upper limit.

We will dedicate a future work to study how the black hole growth portrayed here affects the evolution of the stellar mass of the host galaxy.

%***************************************************************************
\section*{Acknowledgements} 
%***************************************************************************

We thank the anonymous referee for helping us improve the clarity of the paper. OP and PD acknowledge support from the European Commission's and University of Groningen's CO-FUND Rosalind Franklin program. PD also acknowledges support from the European Research Council's starting grant ERC StG-717001 (``DELPHI") and from the NWO grant 016.VIDI.189.162 (``ODIN"). TRC acknowledges the support of the Department of Atomic Energy, Government of India, under project no. 12-R\&D-TFR-5.02-0700.

%**************************************************************************
%%%%%%%%%%%%%%%%%%%%%%%%%%%%%%
\bibliographystyle{mn2e}
\bibliography{lib}

\label{lastpage} 
\end{document}